\documentclass[12pt]{article}
\usepackage{graphicx}
\def\Journal#1#2#3#4{, {#1} {\bf #2}, #3 (#4).}
\def\CQG{\em Class.Quantum Grav.}
\def\GRG{\em Gen. Rel. Grav.}

\def\PLA{{\em Physics Letters}  A}

\def\PLA{{\em Phys. Lett.}  A}

\def\PRL{\em Phys. Rev. Lett.}

\def\RMP{\em Rev. Mod. Phys.}

\begin{document}
\begin{center}{\large\bf Rotational Analog of the Hall Effect:\\
Coriolis Contribution to Electric Current}\\ \vspace*{1cm}
\vspace*{9 mm} {\bf B.J. Ahmedov$^{1,2,3}$ and M.J.
Ermamatov$^{1,3}$}\\

\vspace*{9 mm}

{\it $^{1}$Institute of Nuclear Physics and Ulugh Beg Astronomical
Institute\\ Astronomicheskaya 33,
    Tashkent 700052, Uzbekistan\\}
    {\it $^{2}$International Center for
Relativistic Astrophysics, Pescara, Italy}\\ {\it $^{3}$
Inter-University Centre for Astronomy and Astrophysics\\
         Post Bag 4 Ganeshkhind,
411007 Pune, India}
\end{center}

\vspace*{20mm}

{\small A galvanogyroscopic effect which is the rotational
analog of the gravitomagnetic Hall effect has been proposed.
As a consequence of Ohm's law in the rotating frame, the effect of the
Coriolis force on the conduction current is predicted to give rise to
an azimuthal potential difference $V_{gg}$ about $10^{-3}V$ in a spinning
rotor carrying radial electric current $i_r$.
The potential difference developed by the galvanogyroscopic effect is
proportional both to angular velocity ${\mathbf \Omega}$ and to the electric
current.}

\newpage

Our aim here is to incorporate a Coriolis force term to the theory of
electrical conductivity, which could lead to a galvanogyroscopic effect.
Then the predicted effect is applied to design a verifiable experiment
to detect an influence of the rotation of the earth
on an electric current flowing in a conductor.

The point is that in the rotating frame of reference
conduction electrons are affected by the Coriolis force
\begin{eqnarray}
{\mathbf F}=2m[{\mathbf v}\times{\mathbf \Omega}]\ ,
\end{eqnarray}
${\mathbf v}$ is the velocity of the conduction
electrons, $m$ is the electron mass and
${\mathbf \Omega}$ is the angular velocity of rotation.

According to the Larmor's theorem the Coriolis force imitates
the equivalent magnetic field ${\mathbf B}_{eq}$ applied to the immovable
medium:

\begin{eqnarray}
{\mathbf F}=-(|e|/c)[{\mathbf v}\times{\mathbf B}_{eq}]\ ,
\quad {\mathbf
B}_{eq}=-(2mc/|e|){\mathbf \Omega}\ ,
\label{eq:larmor}
\end{eqnarray}
where $e=-|e|$ is
the electronic charge and $c$ is the velocity of light in vacuum.  That is
the
effect of the Coriolis force on conduction electron is the same as that of
the
Lorentz force due to the equivalent magnetic field.

In the Hall effect$^{(1)}$, under influence of a magnetic
field on conduction current, an electric field
\begin{eqnarray}
{\mathbf E}_H=R_H{\mathbf j}\times{\mathbf B}
\label{eq:hall}
\end{eqnarray}
appears in a conductor to which an electric
current ${\mathbf j}$ and a magnetic field ${\mathbf B}$
are applied in perpendicular directions, where $R_H$ is the Hall constant.
Consequently according to equations equations (\ref{eq:larmor}) and
(\ref{eq:hall}) there exists a galvanogyroscopic effect which can be summarized
by saying that an inner electric field being perpendicular to the current and
angular velocity is induced by rotation in a spinning crosslike rotor carrying
a radial electric current $i_r$ (see Figure).  Then the voltmeter applied in
an azimuthal direction would detect the
galvanogyroscopic voltage $V_{gg}$. This
effect has the following properties: (i) galvanogyroscopic effect is odd, that
is, the galvanogyroscopic voltage changes sign in dependence of the polarity of
either the angular velocity or the electric current, (ii) in the first
approximation, the voltage $V_{gg}$ is proportional to $\Omega$ and described
by the formula
\begin{equation}
V_{gg}=(R_{gg}/d)i_r\Omega\ , \
\label{eq:gg}
\end{equation}
where $R_{gg}={2m}/{ne^2}$ is the galvanogyroscopic coefficient,
 $n$ is the electron concentration, and $d$ is the
thickness of the plate of the rotor.

\begin{figure}
\begin{center} \label{fig1}
\includegraphics[height=60mm, width=86mm]{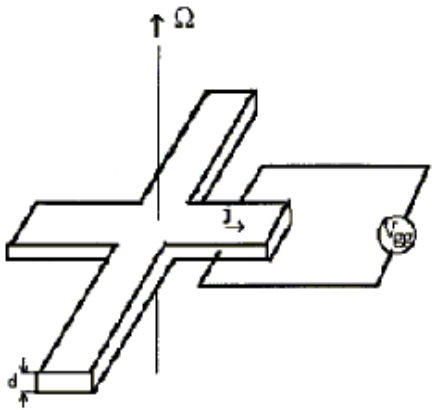}
\end{center}
\caption{Crosslike rotor carrying radial electric current $i_r$
where the galvanogyroscopic voltage $V_{gg}$ is produced. }
\end{figure}

The galvanogyroscopic effect may be regarded, among other things, as a
consequence of Ohm's law in the rotating frame which can be obtained
by extending the usual derivation of
constitutive equations to include noninertial acceleration. The
equation of motion for conduction electrons
in rotating frame can be written as
\begin{equation}
m\frac{d{\mathbf v}}{dt}-m{\mathbf \Omega}\times({\mathbf r}
\times{\mathbf \Omega})-2m{\mathbf\Omega}=-e({\mathbf E}+
\frac{1}{c}{\mathbf v}\times{\mathbf B})+
\frac{ne^2}{\sigma}{\mathbf v}\ ,
\label{em1}
\end{equation}
where the last term is due to the resistance force acting on conduction
electrons from continuous medium with conductivity $\sigma$,
${\mathbf E}$ is the electric field.
It is clear that the equations (\ref{em1}) would involve the effective
mass of the conduction electrons, rather than the free mass. However
as it was discussed and demonstrated by many authors  (see$^{(2-4)}$
for review and most recent results) that the use of the free mass of the
electron instead of the effective one is very good approximation on
measurement of electromotive force in accelerated conductors.

Taking into account the fact that the steady state
$\frac{d{\mathbf v}}{dt}=0$ establishes within the conductor
in extremly short time $(\tau\approx{10^{-13}}s)$, one arrives
at Ohm's law for the current density ${\mathbf j}=ne{\mathbf v}$
which flows in the rotating conductor
\begin{equation}
{\mathbf E}=\frac{1}{\sigma}{\mathbf j}+R_H{{\mathbf j}\times
{\mathbf B}-
R_{gg} {\mathbf j}\times{\mathbf \Omega}+
\frac{m-\gamma M_a}{e}{\mathbf \Omega}\times
({\mathbf r}\times\mathbf \Omega})\ .
\label{eq:rotohm}
\end{equation}

Except for ordinary standard terms, the current flowing in the rotating
conductor has two contributions: one is due to
the Coriolis force effect on current which is represented by the term
$R_{gg}{\mathbf j}\times{\mathbf \Omega}$
and other one is due to the centrifugal force. The correction
$\gamma M_a$ ($\gamma$ is the parameter of order $1$ and $M_a$ is the
atomic mass)
in the right hand side of the equation (\ref{eq:rotohm}) is due to
the strain gradients produced
by the acceleration field (see, for example,$^{(5)}$).

Ohm's law for conduction current has, moreover, been generalized
to include effects of gravity and inertia in recent
papers$^{(4-10)}$ where, in particular, the electromagnetic fields
arising from the centrifugal acceleration in the rotating
conductors are calculated. However the effect of the Coriolis
force upon an electric current has been taken into account only
in$^{(9)}$ where the experimentally measured magnetic
field$^{(10)}$ around a rotating conductor was explained as a
result of azimuthal current produced by the effect of Coriolis
force on radial thermoelectric current. Furthermore it has been
experimentally established that centrifugal forces produce a
potential drop in a spinning conductor$^{(4,12)}$ (see$^{(3)}$ for
review) and consequently act on the electrons within accelerating
conductors as a true electric field applied to the conductor.

Consider now a crosslike rotor carrying radial current and spinning with a
high angular velocity as in Figure. Suppose that the external magnetic
field is absent.
Charges moving radially inward or outward will experience a Coriolis
force\footnote{The centrifugal force being proportional to $\Omega^2$
has been neglected in the linear approximation in the
angular velocity of the rotation ${\mathbf \Omega}$, which
is hereafter used for simplicity of calculations.}, which
will cause some motion of charges relative to the lattice
in the azimuthal direction. This motion will be compensated by an azimuthal
electric field arising from charge separation on opposite sides of rotor.
As we can see from Ohm's law (\ref{eq:rotohm}), the effect of the Coriolis
force on the current gives rise to an electric field perpendicular to the
current,
whose magnitude is defined by (\ref{eq:gg}).

The potential difference produced should be
minute in metals, but appreciable in semiconductors because
the concentration of electric carriers $n\approx 10^{-14}cm^3$
much smaller in semiconductors in comparision to that in
conductors. The sign (+ or -) of the coefficient $R_{gg}\approx 0.8\times
10^{-22}s$ depends on the electric charge of the majority carriers,
hence measurement of the $R_{gg}$ coefficient should determine
both the sign and the concentration of the majority carriers.

For typical values of parameters $\Omega=10^3s^{-1}$, $i_r=10^3A$
and $d=10^{-2}cm$ in the absence of any additional effects one
would expect that $V_{gg}=0.7\times 10^{-2}V$. An experiment to
observe this voltage is to rotate a rotor carrying radial current
and measure the potential difference through it in azimuthal
direction (for example, radial current in the rotor can be
supported by thermoelectric effects as in the
experiment$^{(11)}$). The most dangerous sources of experimental
error, arising from the earth's magnetic field, ionization in
surrounding air etc. should be circumvented during the experiment
with the required accuracy. Measuring the galvanogyroscopic
potential difference as a function of conduction current thus
yields a direct measure of rotation rate which, in principle, can
be used for construction of new type devices such as high
sensitivity speedometers of angular velocity.

If the semiconductor carrying radial current is at rest with
respect to the earth, then there would be a voltage due to the
earth's diurnal rotation with angular velocity
$\Omega_\oplus\approx7.3\times 10^{-5}s^{-1}$. From real
estimations, for parameters $i_r=10^3A$ and $d=10^{-2}cm$ one
finds that an azimuthal voltage $V_\oplus\approx 0.5\times
10^{-10}V$ will be produced by galvanogyroscopic effect across a
semiconductor rigidly rotating together with the earth. The
measurement of this voltage based on advanced experimental
techniques$^{(13)}$ should reprove the diurnal rotation of the
earth, and in this sense, is analogous to the mechanical
measurements made with the Foucault pendulum$^{(14)}$ for
observation of earth's rotation.

In this paper, we have shown that the effect of the Coriolis
force on the conduction current is to induce a galvanogyroscopic
potential difference. Thus electrodynamic analog of the well known
mechanical Beer's law arising from the effect of the Coriolis force
on the stream of a river has been found. Similarly, according to
the galvanogyroscopic effect one lateral side of the rotating conductor
with electric current has opposite electric charges relative to the
other one due to the interaction between electric current
and Coriolis force.

\section*{ACKNOWLEDGMENTS}

BA and ME thank the IUCAA for warm hospitality during their stay
in
 Pune, AS-ICTP and TWAS for the travel support. This research is
also supported in part by the UzFFR (project 01-06) and projects
F.2.1.09, F2.2.06 and A13-226 of the UzCST. BA acknowledges the
partial financial support from NATO through the reintegration
grant EAP.RIG.981259. The authors thank Professor Larry Horwitz
and anonymous referees for their useful comments.

\newpage
{\bf REFERENCES}\\
\begin{enumerate}
\item
        E.H. Hall, ``On a new action of the magnet on
electric currents"\Journal{\em Am. J. Math.}{2}{287-292}
{1951}
\item
      T.J. Davis and G.I. Opat,``Electric fields in
accelerating conductors"\Journal{\CQG}{5}{1011-1028}{1988}
\item
      T.W. Darling, F. Rossi, G.I. Opat, and
G.F. Moorhead
``The fall of charged particles
under gravity: A study of expremental
problems"\Journal{\RMP}{64}{237-257}{1992}
\item
        G.F. Moorhead and G.I. Opat,``Electric fields in accelerating
conductors: measurement of the EMF in rotationally
accelerating coils"\Journal{\CQG}{13}{3129-3139}{1996}
\item
      B.J. Ahmedov and L.Ya. Arifov,
``Principles for detecting charge
redistribution produced by fields of gravity and inertia
inside conductors"\Journal{\GRG}{26}{1187-1195}{1994}
\item
       J. Anandan,
``Relativistic thermoelectromagnetic gravitational
    effects in normal conductors and
superconductors"\Journal{\PLA}{105}{280-284}{1984}
\item
       J. Anandan,
``New relativistic gravitational
effects using charged-particle
interferometry"\Journal{\GRG}{16}{33-41}{1984}
\item
O. Gron,``Application of
Shiff's rotating frame
electrodynamics"\Journal{\em Int. J. Theor. Phys.}{23}{441-448}{1984}
\item
B.J. Ahmedov, ``General relativistic Ohm's law and Coriolis force
effects in rotating conductors"\Journal{\em Gravit.
Cosmology}{4}{139-141}{1998}
\item
B.J. Ahmedov and M.J. Ermamatov, ``Electrical conductivity in
general relativity"\Journal{\em Found. Phys. Lett.}{15}{137-151}
{2002}
\item
B.V. Vasil'ev, ``Thermogyromagnetic
effects"\Journal{\em Russian Physics J.E.T.P.
Letters}{60}{47-50}{1994}
\item
      J.W. Beams, ``Potentials on rotor
surfaces"\Journal{\PRL}{21}{1093-1096}{1968}
\item
A.K. Jain, J.E. Lukens, and J.S. Tsai, ``Test for relativistic gravitational
         effects on charged particles"\Journal{\PRL}{58}{1165-1168}{1987}
\item
      L. Foucault,  {\em Travaux Scientifiques de Foucault}
(Paris, 1878).
\end{enumerate}

\end{document}